\documentclass[a4paper,11pt]{article}
\usepackage{pos}
\usepackage{graphicx}
\graphicspath{{images/}}
\usepackage{parskip}
\usepackage{fancyhdr}
\usepackage{subcaption}
\usepackage{float}
\usepackage{wrapfig}

\title{A next-generation optical sensor for IceCube-Gen2}
 \ShortTitle{IceCube Gen2 LOM}

\author{The IceCube-Gen2 Collaboration \\{\normalsize \normalfont(a complete list of authors can be found at the end of the proceedings)}}
\emailAdd{vbasu@wisc.edu}
\emailAdd{aya@hepburn.s.chiba-u.ac.jp}
\emailAdd{m.dittmer@uni-muenster.de}
\emailAdd{shimizu@hepburn.s.chiba-u.ac.jp}  

\abstract{For the in-ice component of the next generation neutrino observatory at the South Pole, IceCube-Gen2, a new sensor module is being developed, which is an evolution of the D-Egg and mDOM sensors developed for the IceCube Upgrade. The sensor design features up to 18 4-inch PMTs distributed homogeneously in a borosilicate glass pressure vessel. Challenges arise for the mechanical design from the tight constraints on the bore hole diameter (which will be 2 inches smaller than for IceCube Upgrade) and from the close packing of the PMTs. The electronics design must meet the space constraints posed by the mechanical design as well as the power consumption and cost considerations from over 10,000 optical modules being deployed. This contribution presents forward-looking solutions to these design considerations. Prototype modules will be installed and integrated in the IceCube Upgrade.

\vspace{4mm}
{\bfseries Corresponding authors:}
Vedant Basu$^{1*}$, Aya Ishihara$^{2}$, Markus Dittmer$^{3}$, Nobuhiro Shimizu$^{2}$\\
{$^{1}$ \itshape WIPAC, UW Madison, WI 53706, USA}\\
{$^{2}$ \itshape ICEHAP, Chiba University, Chiba, Japan}\\
{$^{3}$ \itshape Institut für Kernphysik, Westfälische Wilhelms-Universität Münster}\\[4mm]
$^*$ Presenter
\FullConference{37$^{\rm{th}}$ International Cosmic Ray Conference (ICRC 2021)\\
		July 12th -- 23rd, 2021\\
		Online -- Berlin, Germany}

}

\begin{document}
\maketitle
\section{From Upgrade to Gen2}
% \vspace{-0.2cm}
The IceCube Neutrino Observatory is a cubic-kilometer neutrino detector constructed at the South Pole \cite{aartsen2017icecube}. Reconstruction of the incident neutrinos
 relies on the detection of Cherenkov photons emitted by charged particles produced via neutrino interactions in the surrounding ice or  bedrock. The photons are collected by Digital Optical Modules (DOM), installed within holes drilled into the ice. \textbf{IceCube-Gen2}, a planned expansion of IceCube, aims to increase the rate of observed cosmic neutrinos tenfold compared to IceCube, and to be able to detect sources five times fainter. %The \textbf{IceCube Upgrade} \cite{Ishihara2019:IC_Upgrade}, as a first step, will enhance IceCube's capabilities with new optical modules on seven strings in the DeepCore region.
\vspace{-0.5cm}
\subsection{Upgrade DOMs}
\begin{figure}[h!]
	\begin{minipage}{0.45\linewidth}
	\vspace{-1.2cm}
		\centering
	\includegraphics[width=0.6\linewidth]{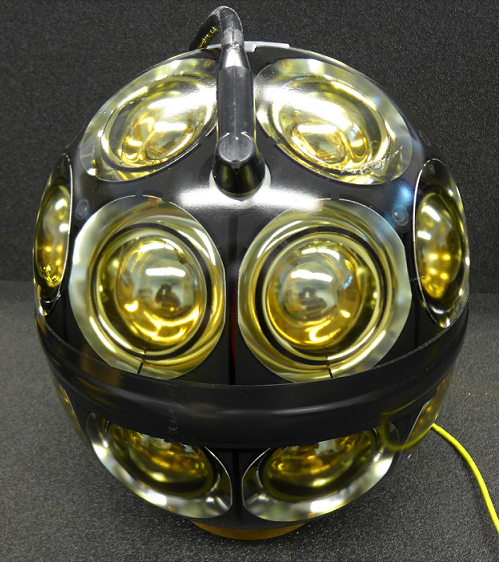}
	\caption{The multi-PMT Digital Optical Module (mDOM).}
	\label{fig:mdompic}
\end{minipage}
\hspace{0.05\linewidth}
\begin{minipage}{0.45\linewidth}
\vspace{-0.5cm}
		\centering
	\includegraphics[width=0.5\linewidth]{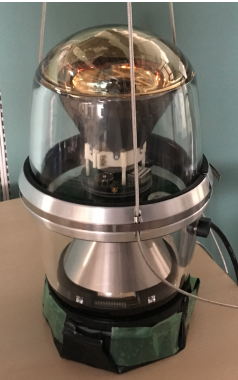}
	\caption{The  dual-PMT Optical Module (D-Egg).}
	\label{fig:D-Eggpic}
\end{minipage}
\end{figure}

\vspace{-0.3cm}
The IceCube Upgrade \cite{Ishihara2019:IC_Upgrade}, as a first step, will consist of nearly 700 new optical modules, as an extension of IceCube to lower energies and a testbed for Gen2 modules. One  design for IceCube Upgrade optical modules, the mDOM (Figure~\ref{fig:mdompic}), features 24 Photomultiplier Tubes (PMTs) of 3" diameter, yielding an almost homogeneous angular
coverage and providing an effective photosensitive area more than twice that of the Gen1 IceCube DOMs. Another module for the Upgrade is 
the D-Egg (Figure~\ref{fig:D-Eggpic}), which has two 8" PMTs facing opposing directions.
\vspace{-1.5cm}
\subsection{Gen2 Module Design}
\vspace{-0.4cm}
The Long Optical Module (LOM) is an evolution of the modules used in the IceCube Upgrade. The sensor diameter has been minimized to enable the module to fit within a narrower hole, saving time and fuel costs associated with hole-drilling. Multiple PMTs have been used to maximize effective area per sensor while maintaining multi-directional sensitivity. The electronics have been optimized for a reduced power consumption of 4W per module, and easier manufacture and testing.\\
With these goals in mind, the LOM will use new 4" diameter PMTs to obtain the same effective area as the mDOM with fewer channels, reducing the power consumption due to digitization and processing. The waveform processing has been shifted to the PMT base, integrating digitization and high voltage (HV) generation and moving from a monolithic central processing architecture to a modular framework.
In addition, the vessel diameter has been limited to $\approx$12". For reference, the Gen1 DOMs were 13" in diameter, while the mDOM is 14" wide, and we save up to 10\%  for every inch reduction in hole diameter. This correlation breaks below a drill diameter of 30cm, and therefore a design narrower than the one selected does not offer measurable benefits for drilling.\\
\begin{minipage}{0.4\textwidth}
	\begin{figure}[H]
		\centering
		\includegraphics[width=\linewidth]{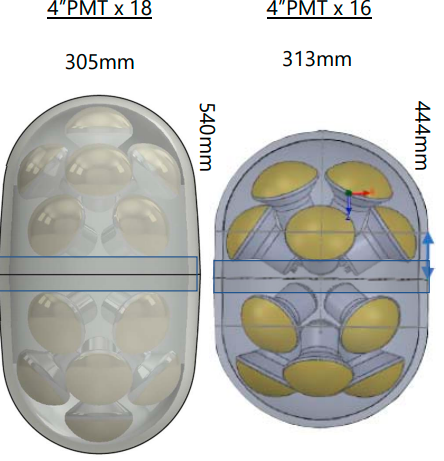}
		\caption{The two prototype designs.}
		\label{fig:LOMs}
	\end{figure}
\end{minipage} \hfill
\begin{minipage}{0.55\textwidth}
\vspace{-1.5cm}
The pressure vessel designs are a major factor in the size of the LOM, and the number of PMTs which can fit. Two designs are currently under development (Fig.~\ref{fig:LOMs}), with 18 and 16 PMTs. Simulation results yield an effective area (400nm) of $118~\rm{cm}^2$ for the 18-PMT module, and $105~\rm{cm}^2$ for the 16-PMT module, an increase in Cherenkov-averaged effective area of 3.2-4.2 times the Gen-1 DOM using PMTs with 25\% quantum efficiency\cite{Shimizu:2021Simulation}.
In the near term, the goal is to build ten modules of each type, with a total of 12 R\&D modules within the Upgrade deployments. These will be further developed into a single design for Gen2, which will have over 10000 modules distributed across 120 strings \cite{aartsen2021icecube}. 
\end{minipage}
\vspace{-0.3cm}
\section{Mechanical Structure}
The primary requirements for the mechanical support structure are to hold the PMTs in the correct orientation, while absorbing the compressive shrinkage of the vessel.  Notwithstanding the constraints on vessel diameter, a minimum clearance between components of 4-5mm is maintained, requiring the support structure to push components radially outward away from each other. The support structure design also dovetails closely with the potting procedure to interface the PMTs with the vessel. 
\subsection{Optical Coupling of PMTs}
Each PMT is coupled to the wall of the pressure vessel using transparent silicone gel. This could be injected directly into the cavity between a 'shell' and the vessel, to cure in place once installed.
\begin{minipage}{0.3\textwidth}
	\begin{figure}[H]
	\vspace{1.5cm}
	\includegraphics[width=1\linewidth]{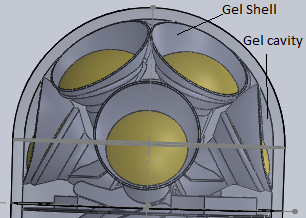}
	\caption{Optical Gel.}
	\label{fig:OpticalGel}
	\end{figure}
\end{minipage} 
\begin{minipage}{0.3\textwidth}
	\begin{figure}[H]
	\centering
	\includegraphics[width=\linewidth]{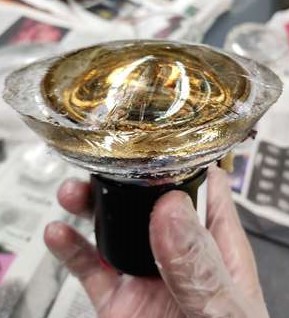}
	\caption{Gel Pads.}
	\label{fig:GelPad}
	\end{figure}
\end{minipage}
\begin{minipage}{0.37\textwidth}
\vspace{0.2cm}
	 The mDOM uses a 3D printed support structure, with reflectors to detect any photons incident off-axis. For the LOM, an alternate approach is taken, using \textbf{gel pads} moulded to fit between the PMTs and the vessels. Total Internal Reflection (TIR) at the conical surface replaces the opaque reflectors with comparable efficiency. The gel pad must be stiff enough to maintain shape while pliable enough to cushion forces without damaging the PMT. GEANT simulations are used to determine the optimal shape for the gel pads ~\cite{Shimizu:2021Simulation}.
\end{minipage}
\vspace{2mm}
\\A major obstacle in the gel pad approach is the occurrence of bubbles, which act as photon scattering centres. These bubbles expand under vacuum, which is problematic as both module halves are sealed using under-pressure.
The optical gel is a two-part addition-curing silicone, requiring three days to cure. The gel is first degassed to obviate the bubbles introduced during the mixing process, a process referred to as \textbf{primary degassing}.
The gel is then poured into a mould for gel pad casting, directly onto the face of the PMT. 
%We find that siphoning, using an elevated funnel and a syringe to pressurize the tube, introduces very few bubbles, along with allowing the gel to flow along the walls of the receptacle as opposed to pouring it in a stream. 
%We use Multi-Jet Fusion 3D printed moulds for current testing, with a 1:1 mix of dish soap and water as a mould release agent. \\
Once cast, the PMT and its gel pad assembly must be coupled to the vessel wall without any bubbles at the interface. Initial tests involved a layer of uncured gel at the surface of the pad, which was then compressed against the vessel. The method failed for PMTs angled to the vertical, as the liquid gel leaked out leaving air cavities behind.\\
\begin{minipage}{0.35\textwidth}
	\vspace{2mm}
An attempt to solve the issue included casting a pad with a flat surface, creating a cavity with the curved pressure vessel. Silicone caulk is used to seal the rim of the pad to the pressure vessel, and the cavity is then filled with liquid gel. mDOM testing indicates that the gel must be degassed after pouring (\textbf{secondary degassing}), to mitigate the risk of delamination at low temperature and vacuum conditions.
\end{minipage}\hfill
\begin{minipage}{0.3\textwidth}
\vspace{-1.2cm}
	\begin{figure}[H]
		\centering
		\includegraphics[width=1\linewidth]{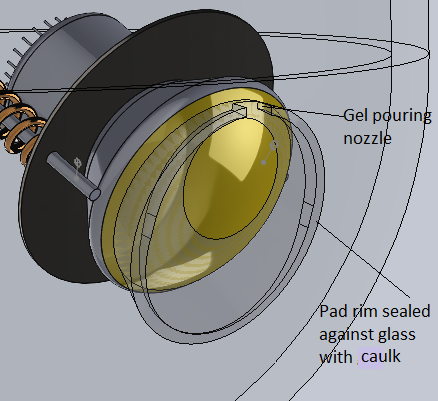}
		\caption{ Cavity method.}
		\label{fig:cavity method}
	\end{figure}
\end{minipage} 
\begin{minipage}{0.3\textwidth}
\vspace{-1.15cm}
	\begin{figure}[H]
		\centering
		\includegraphics[width=\linewidth]{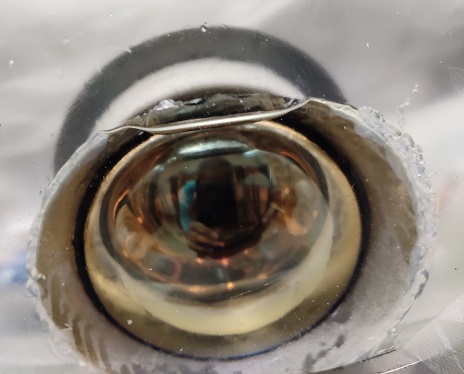}
		\caption{Removal of bubbles after secondary degassing.}
		\label{fig:CavityPad}
	\end{figure}
\end{minipage}
\begin{minipage}{0.4\textwidth}
	\vspace{-0.2cm}
	\begin{figure}[H]
		\centering
		\includegraphics[width=0.6\linewidth]{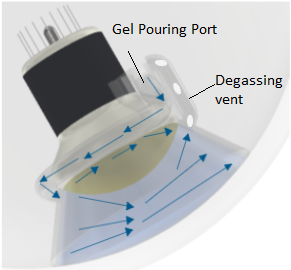}
		\caption{A transparent gel 'shell'.}
		\label{fig:GelShell}
	\end{figure}
\end{minipage} 
\begin{minipage}{0.6\textwidth}
		\vspace{-1cm}
An alternative approach under consideration is the use of transparent plastic shells, to be sealed in with the PMTs. The gel is poured into the shell to cure. The shell material must be selected to match the coefficient of thermal expansion of the gel, allowing it to shrink at low temperatures without delaminating from the gel. It must also retain a high degree of transparency, while being economical to manufacture at large scale.
\end{minipage}
% \vspace{-3cm}
\subsection{Support Structure}
\begin{minipage}{0.48\textwidth}
% 	\vspace{-1cm}
	The support structure must be designed keeping in mind the requirements of the PMT interfacing procedure. In Figure~\ref{fig:FacetedSupport}, a structure welded together out of stainless steel sheets is shown. The PMTs are aligned using plastic guide bearings, with a foam collar. To exert the necessary outward pressure for proper gel pad interfacing, an inflatable bladder is proposed around the neck of the PMT, which permits articulation during assembly. These features combine to yield an economical solution scalable for mass production.
\end{minipage}\hfill
\begin{minipage}{0.5\textwidth}
		\vspace{-2cm}
	\begin{figure}[H]
		\includegraphics[width=1\linewidth]{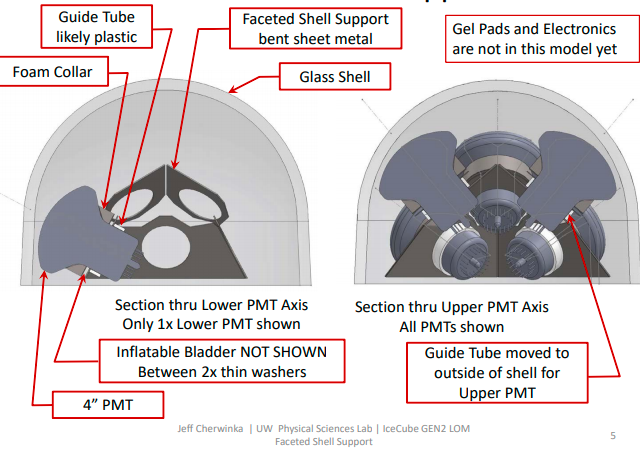}
		\caption{Faceted support structure, shown for the 16 PMT configuration.}
		\label{fig:FacetedSupport}
	\end{figure}
\end{minipage} 
\clearpage
\begin{minipage}{0.35\textwidth}
		\centering
		\includegraphics[width=0.8\linewidth]{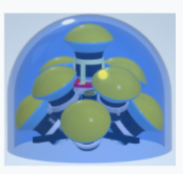}
		\captionof{figure}{Tree Shaped Support, shown for the 18 PMT configuration.}
		\label{fig:TreeSupport_1}
\end{minipage}
% \begin{minipage}{0.35\textwidth}
% 	\vspace{-0.5cm}
% 		\centering
% 		\includegraphics[width=1.1\linewidth]{images/TreeShapeSupport}
% 		\captionof{figure}{}
% 		\label{fig:TreeSupport}
% \end{minipage}\hfill
\begin{minipage}{0.65\textwidth}
% 		\vspace{-2cm}
An alternative design is shown in Figures ~\ref{fig:TreeSupport_1}, where each PMT is aligned with a flexible plastic collar, connected to a central shaft. Springs are used to provide the required outward pressure.
\section{Electronics}
	\vspace{-0.4cm}
A prime motivation in using larger PMTs for the LOM is to obtain a photon effective area comparable to the mDOM with  fewer channels, reducing power consumption. The electronics design for the LOM builds upon the system used in the mDOM, expanding the mDOM MicroBase, which regulates the PMT HV generation, to include data acquisition (DAQ) functionality. The new bases are referred to as the \textbf{Waveform MicroBase}.The digitization and processing have thus been distributed to the base of each PMT, moving to a modular design for ease of assembly and manufacturing.
\end{minipage}
% \vspace{-0.4cm}
% \begin{minipage}{0.4\textwidth}
% % 	\vspace{-0.2cm}
% % 	\begin{figure}[H]
% 			\centering
% 		\includegraphics[width=0.8\textwidth]{images/Polar_Microbase}
% 		\captionof{figure}{An illustration of the new base shapes.}
% \label{fig:Eq_Microbase}
% % 	\end{figure}
% \end{minipage}\hfill
% \begin{minipage}{0.58\textwidth}
% 		\vspace{0.3cm}
  To maximise packing efficiency and board surface area while maintaining the necessary clearances between components, each PMT base has a special shape to fit in the gaps between neighbouring PMTs 
% (Fig ~\ref{fig:Eq_Microbase}).
% \end{minipage}
\subsection{Central Boards}
Following the existing IceCube Upgrade mDOM and D-Egg designs, a central processing board will provide common data management and control functions, as well as interfaces to the long in-ice cable connection to "hub" surface computers (Fig.~\ref{fig:WuBaseTopLevelBD}). These parts will match closely the mDOM design described in Ref.~\cite{Classen:2021mDOM}, including an ARM microprocessor (STM32H743 MCU), the Ice Communications Module ("ICM") with a dedicated FPGA, and DC-DC converters supplying logic level power from the higher voltage supply on the in-ice cable. The ICM provides communication and a precise common time base for the module, synchronized to the IceCube master clock.
The LOM has a fanout of communication between the central MCU and the Waveform MicroBase boards, replacing the centralized block for ADCs and associated FPGA, as implemented on prior "Main Board" designs.Ribbon cables carry UART communication, clocks for time synchronization, and low voltage power to the PMT bases.\\
\begin{minipage}{0.25\textwidth}
% 		\vspace{0.2cm}
  The UART speed of $1.5\,{\rm Mbps}$ is able to carry all hit data to the central MCU, using a multiplexer to select bases. To facilitate a multi-level detector trigger and readout scheme~\cite{Kelley:2021DAQ}, a flash memory chip in a circular buffer mode retains all hit data for at least a week.  
\end{minipage}\hfill
\begin{minipage}{0.75\textwidth}
% 	\vspace{0.3cm}
\flushright
% 	\begin{figure}[H]
% 		\centering
		\includegraphics[width=0.97\linewidth]{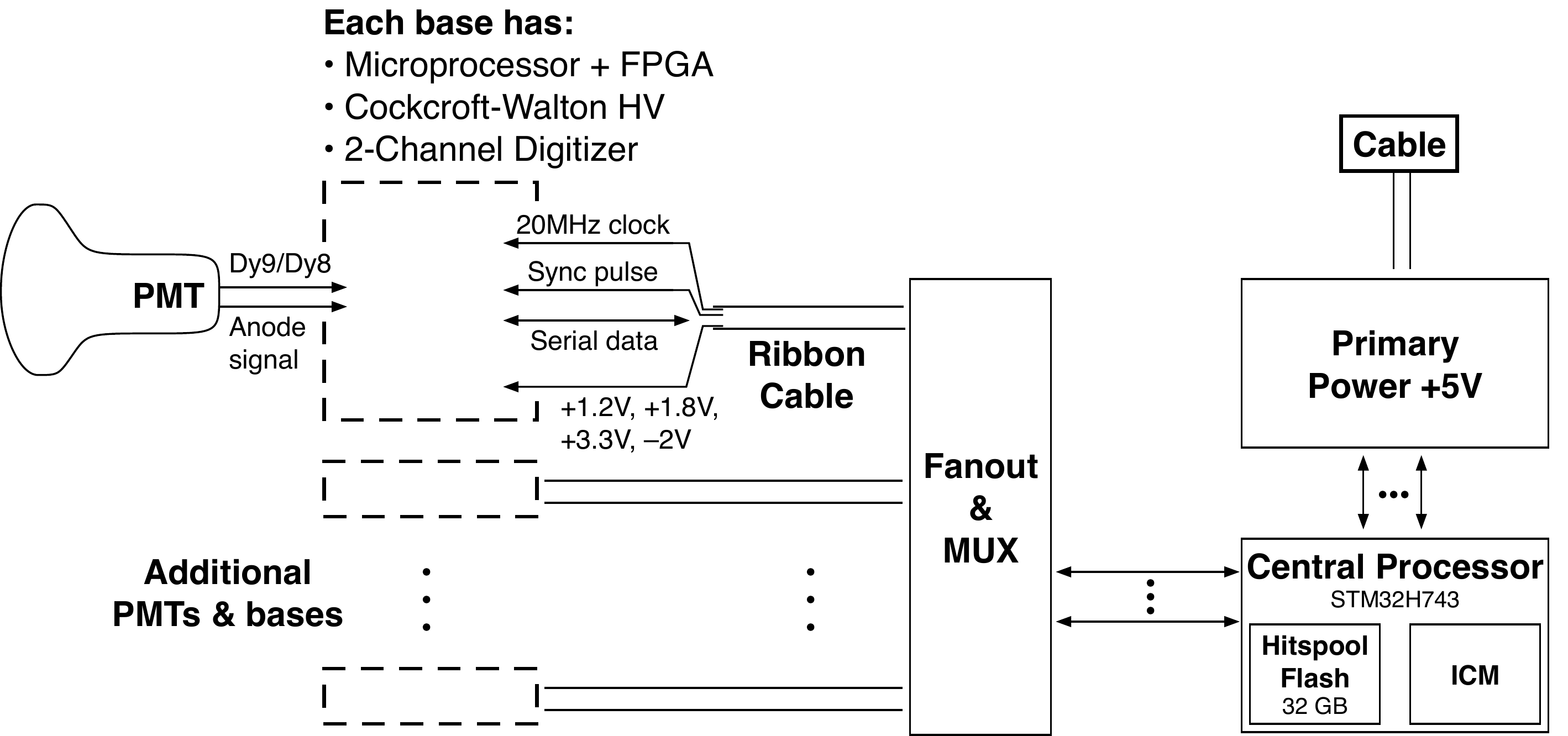}
		\captionof{figure}{The distributed processing architecture.}
		\label{fig:WuBaseTopLevelBD}
\end{minipage}
\clearpage
\subsection{Waveform MicroBase}
\begin{figure}[H]
	\centering
	\includegraphics[width=0.9\linewidth]{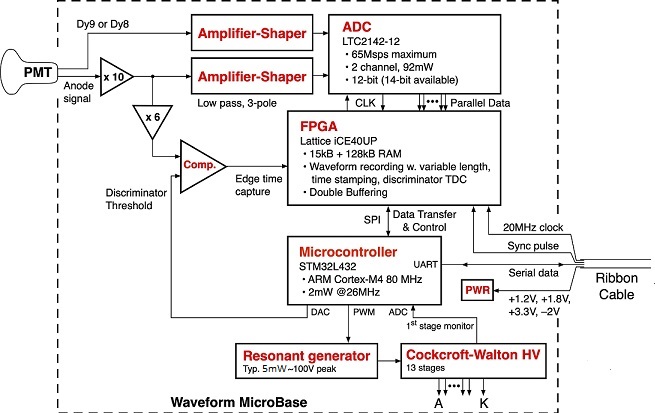}
	\caption{A block diagram of the Waveform MicroBase, demonstrating the integration of DAQ functionality with HV generation at the PMT base.}
	\label{fig:WuBaseBD}
\end{figure}

\begin{figure}[h!]
	%	\centering
	\begin{minipage}[t]{7cm}
		\centering
		\includegraphics[width=1.1\linewidth]{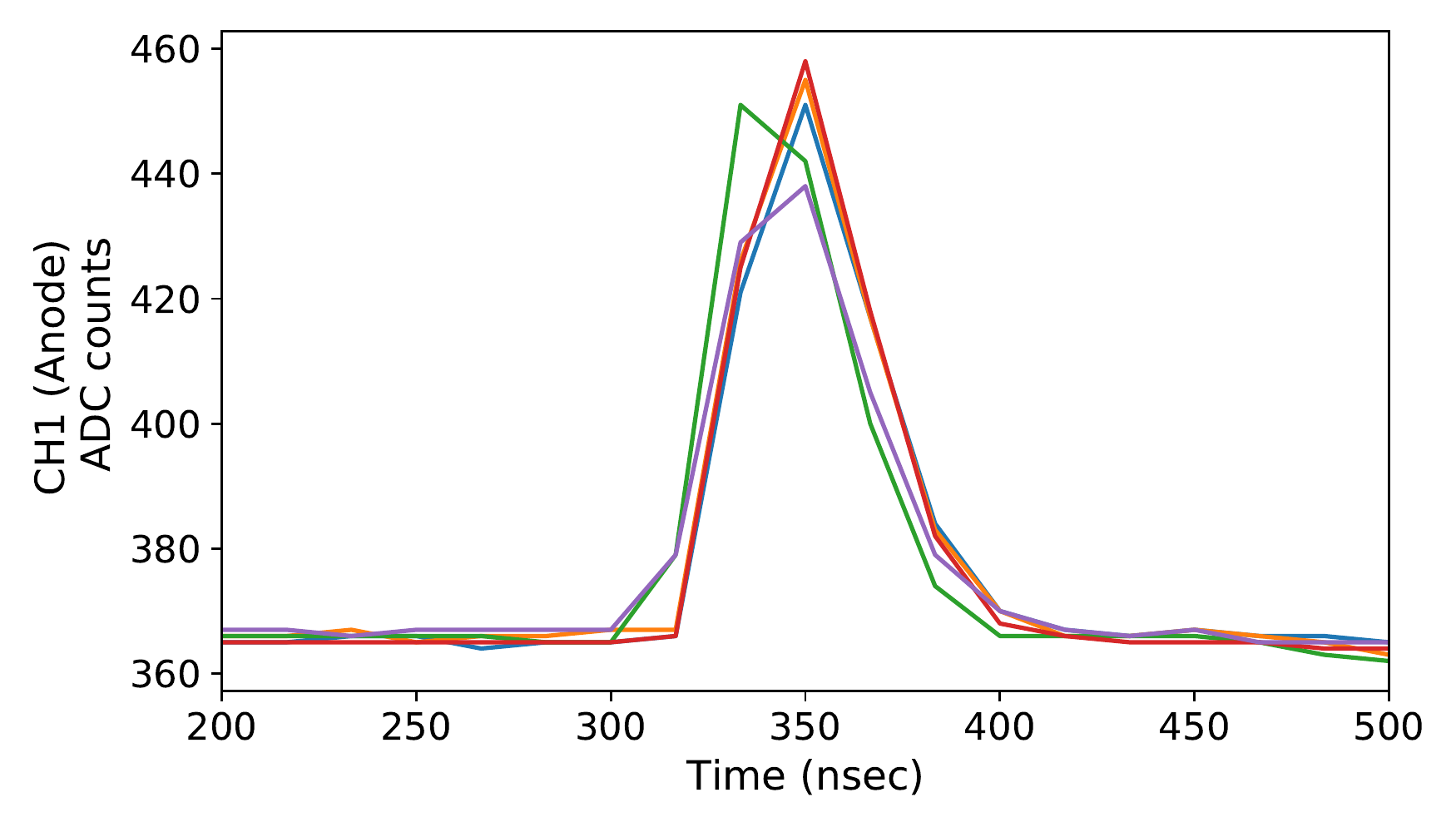}
		\subcaption{}
		\label{fig:SPE_Spectrum_a}
	\end{minipage}
\begin{minipage}[t]{7cm}
	\centering
	\includegraphics[width=0.8\linewidth]{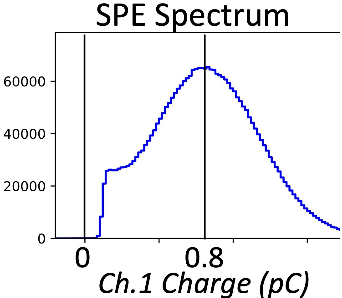}
	\subcaption{}
	\label{fig:SPE_Spectrum_b}
\end{minipage}
	\caption{\textit{Left}: Single Photoelectron waveforms. \textit{Right}: The measured charge distribution for single photoelectrons at a gain of $5\times 10^{6}$.}
	\label{fig:SPE_Spectrum}
\end{figure}

\begin{figure}[h!]
	%	\centering
	\begin{minipage}[t]{7.5cm}
		\centering
		\includegraphics[width=1.1\linewidth]{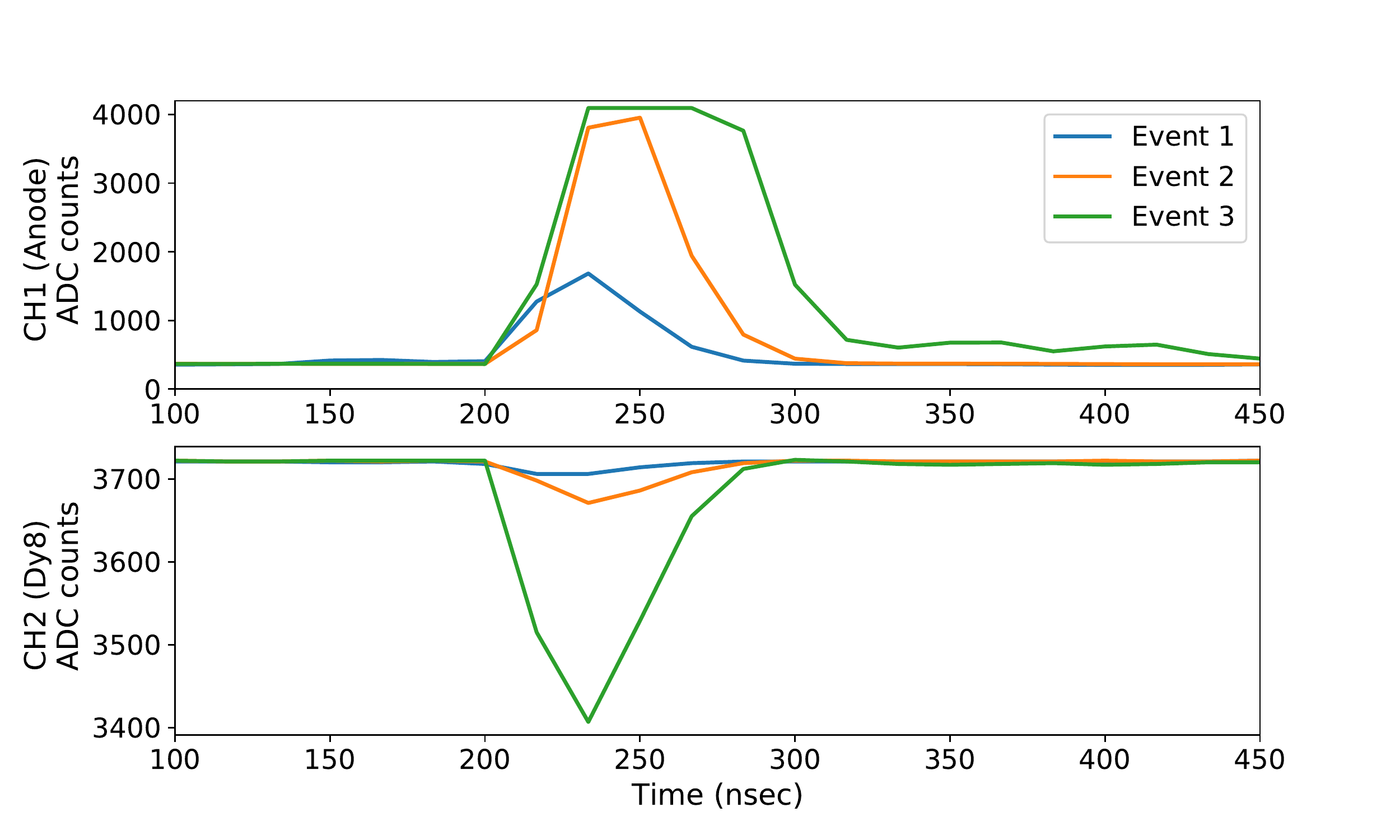}
		\subcaption{}
		\label{fig:Ch1vsCh2_a}
	\end{minipage}
\begin{minipage}[t]{7.5cm}
\centering
		\includegraphics[width=1.1\linewidth]{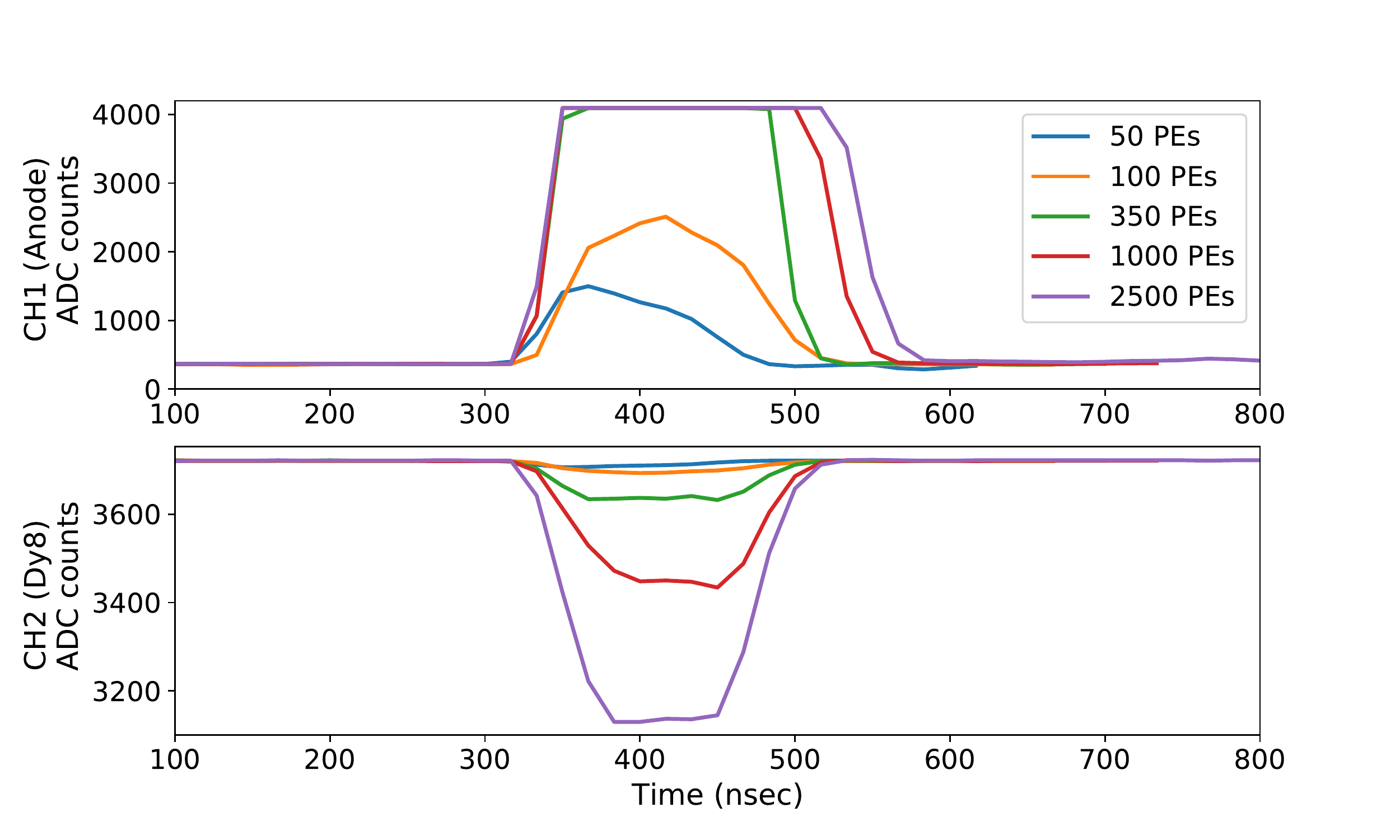}
	\subcaption{}
	\label{fig:Ch1vsCh2_b}
\end{minipage}
	\caption{\textit{Left}: A sample event readout, displaying narrow pulses on both channels. Event 3 has saturated the anode readout on channel 1, but using the readout on channel 2, here taken from dynode 8, one is able to extract information from the event. \textit{Right}: The response to wide pulses. The 350 PE event has saturated the anode readout on channel 1, but channel 2 retains linearity.}
	\label{fig:Ch1vsCh2}
\end{figure}

The Waveform MicroBase (Figure~\ref{fig:WuBaseBD}) has been optimized for low power consumption and wide dynamic range. The high voltage section is a resonant Cockcroft-Walton generator, identical to the well-tested mDOM base~\cite{Classen:2021mDOM}, consuming about $5\,{\rm mW}$. The same software for control and regulation of the high voltage is run on a low-power STM32L4 microcontroller, which now additionally manages buffering and low-level processing of digital waveform data.  \\
IceCube neutrino events yield PMT signals varying widely in intensity, depending both on the neutrino energy and the distance of each module from the light source. The event reconstruction relies on the amount of light received by each PMT, the precise times of arrival, and the detailed arrival time profile. To best capture this information, the base includes two analog channels, one for the PMT anode and another connected to a dynode (Dy8) with lower gain. These two signals are shaped and digitized continuously in a 2-channel ADC at $60\,{\rm MSPS}$, and captured in a low power FPGA. The recorded waveform is a direct representation of the photon intensity and time profile, with individual photons appearing in Channel~1 as shaped pulses $40\,{\rm ns}$ wide and 90~ADC counts high (Figure~\ref{fig:SPE_Spectrum_a}). The anode channel remains linear for multi-photon events with intensities up to $\rm 50\,PE\,/\,25\,nsec$.\\
A discriminator triggers on the rising edge of the first photon and signals the FPGA to store the whole waveform with up to 256 time samples, including pre-trigger and post-trigger intervals. At the same time, a delay line module in the FPGA records the leading edge time with resolution about $1\,{\rm nsec}$. Clock signals on the ribbon cable maintain synchronization with the master clock. 
Figure~\ref{fig:SPE_Spectrum} shows typical single-photoelectron (SPE) waveforms, and the charge distribution from Channel 1, indicating the PMT gain of $5\times 10^6$ and a sharp discriminator threshold set at $0.15\,{\rm SPE}$ with baseline noise of about 1~ADC count. The Channel 1 signal path saturates for large signals, while Channel 2 remains linear up to $\rm 5000\,PE\,/\,25\,nsec$. Figure~\ref{fig:Ch1vsCh2} shows LED flashes of varying brightness and duration, including intensities where both channels are effective and others where Channel 1 saturates at 4095~ADC counts while Channel 2 retains linearity until a higher intensity threshold.

% \subsection{Performance}
% \begin{figure}[h!]
% 	%	\centering
% % 	\begin{minipage}[t]{7cm}
% 		\centering
% 		\includegraphics[width=1\linewidth]{event_sample_twochs_longpulse (1).pdf}
		
% % 	\end{minipage}
% 	\caption{A sample event readout, displaying events on both channels.  }
% 	\label{fig:Ch1vsCh2}
% \end{figure}
% \clearpage
The FPGA uses double buffering methods to avoid deadtime while transferring recorded data to the microcontroller via a SPI link. With expected dark count rates well below $1\,{\rm kHz}$, the link speed ($6\,{\rm Mbps}$) easily manages the data flow. The microcontroller software buffers and prepares waveforms for transfer to the central fanout board. The UART speed of 1.5Mbps is sufficient to carry all hit data blocks from the bases to the central MCU, selecting one base at a time with a multiplexer.
For the larger inter-string spacing of IceCube Gen2, 240m as compared to 125m for Gen 1 and 75m for the Upgrade, most waveforms will have photon time spread of $>25\,{\rm nsec}$, due to scattering of photons traveling tens of meters through ice, and the $60\,{\rm MSPS}$ is sufficient. The chosen sampling rate is a compromise between power consumption and time granularity. The selected ADC is compatible with use of a low power FPGA that consumes only $10\,{\rm mW}$, while the entire base requires $140\,{\rm mW}$, dominated by the continuous ADC operation.

% %\clearpage
\section{Outlook}
An overview of the design and prototyping of new optical modules for IceCube-Gen2 is presented here. The various constraints to be considered and optimized in the LOM are illustrated here, and possible solutions are expounded upon. The guiding principle behind every aspect of the design is to arrive at a procedure that is economical, fault-tolerant and easily scalable for mass production.\\
The design of the mechanical PMT support structure is still evolving in response to the changes in the requirements of PMT interfacing as well as ease of manufacturing. The prototype support structures are being modified to create a feasible assembly procedure.  In the meantime, simulation efforts are in progress to optimize the placement and shape of the gel pads, and to estimate the photon effective area of the two prototype designs. The enhancement of the  photon effective area of a PMT with a gel pad has been experimentally verified.\\
  The design and verification of the electronics systems are proceeding apace, with testing of the Waveform MicroBases underway at various sites. First prototypes are expected to be completed soon and will be ready for deployment as R\&D modules in the 7 string IceCube Upgrade.
%   \section{Acknowledgements}
%   We gratefully acknowledge support from the following agencies and institutions: \textbf{USA}—the US National Science Foundation-Office of Polar Programs, the US National Science Foundation Physics Division; \textbf{Germany}—the Bundesministerium für Bildung und Forschung (BMBF), the Deutsche Forschungsgemeinschaft (DFG), the Helmholtz Alliance for Astroparticle Physics (HAP), the Initiative and Networking Fund of the Helmholtz Association, the Deutsches Elektronen Synchrotron (DESY); \textbf{Japan}—the Japan Society for Promotion of Science (JSPS) and the Institute for Global Prominent Research (IGPR) of Chiba University; 
% \section{Listing some References}\label{sec:refs}

% This is a paper from a previous ICRC \cite{Zoll:2015wcu}. This is a second paper from a previous ICRC \cite{Peiffer:2017vsm}. This is a paper from the current ICRC \cite{Pizzuto:2021icrc}.
% Here is an IceCube journal paper \cite{Aartsen:2016nxy} and an external journal paper \cite{Waxman:1998yy}.

\bibliographystyle{ICRC}
\bibliography{main}

% % \begin{thebibliography}{99}
% % \bibitem{...}
% % ....

% % \end{thebibliography}

% % Full authors list (ONLY FOR COLLABORATIONS)
\clearpage
 \section*{Full Author List: IceCube-Gen2 Collaboration}

% \noindent \textbf{Note comment afterwards:} Collaborations have the possibility to provide an authors list in xml format which will be used while generating the DOI entries making the full authors list searchable in databases like Inspire HEP. For instructions please go to icrc2021.desy.de/proceedings or contact us under icrc2021proc@desy.de.\\

% \scriptsize
% \noindent
% first.author$^1$, 
% second.author$^2$, 
% third.author$^3$ % .... more names
% and 
% last.author$^{n}$ \\

% \noindent
% $^1$first.affiliation.
% $^2$second.affiliation. % .... more affiliation
% $^{m}$last.affiliation.
\scriptsize
\noindent
R. Abbasi$^{17}$,
M. Ackermann$^{71}$,
J. Adams$^{22}$,
J. A. Aguilar$^{12}$,
M. Ahlers$^{26}$,
M. Ahrens$^{60}$,
C. Alispach$^{32}$,
P. Allison$^{24,\: 25}$,
A. A. Alves Jr.$^{35}$,
N. M. Amin$^{50}$
% ...
\\
\\
% ...
$^{12}$ Universit{\'e} Libre de Bruxelles, Science Faculty CP230, B-1050 Brussels, Belgium \\
% ...
$^{17}$ Department of Physics, Loyola University Chicago, Chicago, IL 60660, USA \\
% ...
$^{22}$ Dept. of Physics and Astronomy, University of Canterbury, Private Bag 4800, Christchurch, New Zealand \\
% ...
$^{24}$ Dept. of Astronomy, Ohio State University, Columbus, OH 43210, USA \\
$^{25}$ Dept. of Physics and Center for Cosmology and Astro-Particle Physics, Ohio State University, Columbus, OH 43210, USA \\
$^{26}$ Niels Bohr Institute, University of Copenhagen, DK-2100 Copenhagen, Denmark \\
% ...
$^{32}$ D{\'e}partement de physique nucl{\'e}aire et corpusculaire, Universit{\'e} de Gen{\`e}ve, CH-1211 Gen{\`e}ve, Switzerland \\
% ...
$^{35}$ Karlsruhe Institute of Technology, Institute for Astroparticle Physics, D-76021 Karlsruhe, Germany  \\
% ...
$^{50}$ Bartol Research Institute and Dept. of Physics and Astronomy, University of Delaware, Newark, DE 19716, USA \\
% ...
$^{60}$ Oskar Klein Centre and Dept. of Physics, Stockholm University, SE-10691 Stockholm, Sweden \\
% ...
$^{71}$ DESY, D-15738 Zeuthen, Germany \\
% ...

\subsection*{Acknowledgements}

\noindent
USA {\textendash} U.S. National Science Foundation-Office of Polar Programs,
U.S. National Science Foundation-Physics Division,
U.S. National Science Foundation-EPSCoR,
Wisconsin Alumni Research Foundation,
Center for High Throughput Computing (CHTC) at the University of Wisconsin{\textendash}Madison,
Open Science Grid (OSG),
Extreme Science and Engineering Discovery Environment (XSEDE),
Frontera computing project at the Texas Advanced Computing Center,
U.S. Department of Energy-National Energy Research Scientific Computing Center,
Particle astrophysics research computing center at the University of Maryland,
Institute for Cyber-Enabled Research at Michigan State University,
and Astroparticle physics computational facility at Marquette University;
Belgium {\textendash} Funds for Scientific Research (FRS-FNRS and FWO),
FWO Odysseus and Big Science programmes,
and Belgian Federal Science Policy Office (Belspo);
Germany {\textendash} Bundesministerium f{\"u}r Bildung und Forschung (BMBF),
Deutsche Forschungsgemeinschaft (DFG),
Helmholtz Alliance for Astroparticle Physics (HAP),
Initiative and Networking Fund of the Helmholtz Association,
Deutsches Elektronen Synchrotron (DESY),
and High Performance Computing cluster of the RWTH Aachen;
Sweden {\textendash} Swedish Research Council,
Swedish Polar Research Secretariat,
Swedish National Infrastructure for Computing (SNIC),
and Knut and Alice Wallenberg Foundation;
Australia {\textendash} Australian Research Council;
Canada {\textendash} Natural Sciences and Engineering Research Council of Canada,
Calcul Qu{\'e}bec, Compute Ontario, Canada Foundation for Innovation, WestGrid, and Compute Canada;
Denmark {\textendash} Villum Fonden and Carlsberg Foundation;
New Zealand {\textendash} Marsden Fund;
Japan {\textendash} Japan Society for Promotion of Science (JSPS)
and Institute for Global Prominent Research (IGPR) of Chiba University;
Korea {\textendash} National Research Foundation of Korea (NRF);
Switzerland {\textendash} Swiss National Science Foundation (SNSF);
United Kingdom {\textendash} Department of Physics, University of Oxford.
\end{document}